\documentclass[aps,twocolumn,amsmath,amssymb,superscriptaddress,prb]{revtex4-1}

\usepackage{graphicx}
\usepackage{dcolumn}
\usepackage{subfigure}
\usepackage{bm}

\begin{document}


\title{Density functional theory study of the electronic structure of fluorite Cu$_{2}$Se}

\author{Mikael R{\aa}sander}\email{mikra@kth.se}
\affiliation{%
Department of Materials Science and Engineering, Brinellv. 23 KTH (Royal Institute of Technology), SE-100 44 Stockholm, Sweden
}%
\author{Lars Bergqvist}
\affiliation{%
Department of Materials Science and Engineering, Brinellv. 23 KTH (Royal Institute of Technology), SE-100 44 Stockholm, Sweden
}%
\affiliation{%
SeRC (Swedish e-Science Research Center), KTH, SE-100 44 Stockholm, Sweden}%
\author{Anna Delin}
\affiliation{%
Department of Materials Science and Engineering, Brinellv. 23 KTH (Royal Institute of Technology), SE-100 44 Stockholm, Sweden
}%
\affiliation{%
SeRC (Swedish e-Science Research Center), KTH, SE-100 44 Stockholm, Sweden}%
\affiliation{%
Department of Physics and Astronomy, Uppsala University, Box 516, SE-751 20 Uppsala, Sweden
}%
\date{\today}

\begin{abstract}
We have investigated the electronic structure of fluorite Cu$_{2}$Se using density functional theory calculations within the LDA, PBE and AM05 approximations as well as with the non-local hybrid PBE0 and HSE approximations. We find that Cu$_{2}$Se is a zero gap semiconductor when using either a local or semi-local density functional approximation while the PBE0 functional opens up a gap. For the HSE approximation, we find that the presence of a gap depends on the range separation for the non-local exchange. For the occupied part in the density of states we find that the LDA, PBE, AM05, PBE0 and HSE agree as regards the overall electronic structure. However, the hybrid functionals result in peaks shifted towards lower energy compared to the LDA, PBE and AM05. The valence bands obtained using the hybrid functionals are in good agreement with experimental valence band spectra. We also find that the PBE, PBE0 and HSE approximations give similar results regarding bulk properties, such as lattice constants and bulk modulus. In addition, we have investigated the localization of the Cu d-states and its effect on the band gap in the material using the LDA+$U$ approach. We find that a sufficiently high $U$ indeed opens up a gap but that this $U$ is unphysically large.
\end{abstract}

\maketitle
\section{introduction}\label{sec:intro}
The copper chalcogenides Cu$_{2}$X (X = S, Se, and Te) are of possible technological interest because of their thermo- and photoelectric properties as well as ionic conductivity. Cu$_{2-x}$Se, which is the focus of this study, has in particular received attention due to a high ionic conductivity, with possible applications in solar cells\cite{Okimura1980,Chen1985} as well as a good material for thermoelectric converters.\cite{Liu2012} At room temperature Cu$_{2-x}$Se is a rather good p-type conductor with an optical band gap of 1.23~eV.\cite{Sorokin1966} The Cu$_{2-x}$Se system has a rather complicated atomic structure where the phase diagram consist of two phases: The low temperature $\alpha$-phase and the high temperature $\beta$-phase.\cite{Massalski1990} The $\beta$-phase with the space group Fm$\bar{3}$m has the Se atoms in a face-centered cubic (fcc) environment while the superionic Cu atoms are randomly distributed on interstitial positions in the structure, where the majority of the Cu ions are positioned close to the tetrahedral interstitial sites.\cite{Danilkin2003,Danilkin2011} The low temperature phase is stable up to about 400~K and has a lower symmetry crystal structure, where the Cu atoms are not diffusing. We point out that the $\alpha$ to $\beta$ transition varies with the stoichiometry and for $x=0.15-0.25$ the high temperature phase is stable at room temperature.\cite{Abrikosov1983} In addition, we note that both the high and the low temperature phases can be imagined as deviations from the fluorite crystal structure with the difference lying in the occupation of the tetrahedral interstitial positions.
\par
Recently Liu et al.\cite{Liu2012} showed that Cu$_{2-x}$Se exhibited excellent thermoelectric properties for a bulk material with a thermoelectric figure of merit $zT=1.5$ at 1000~K. The reasons for the favorable thermoelectric properties appear to be a low thermal conductivity due to the quasi-liquid behavior of the superionic Cu atoms combined with a rather high value of the power factor $S^2\sigma$, where $S$ is the Seebeck coefficient and $\sigma$ is the electrical conductivity.\cite{Liu2012} It is interesting that a material with a very simple chemical formula has such a low thermal conductivity since this is usually reported for complex systems containing heavy elements, for example in doped skutterudites and clathrates, or in nanostructures.\cite{Snyder2008} Further investigations of the physical properties of Cu$_{2-x}$Se as well as other similar systems are therefore important in order to obtain a fundamental understanding of the thermoelectric properties of the Cu$_{2-x}$Se system in particular and superionic solids in general. In this paper we will present results obtained by  density functional theory calculations on the electronic structure of Cu$_{2}$Se.
The aim with the present study is to obtain a theoretical understanding of the electronic structure of Cu$_{2}$Se by means of density functional theory (DFT) calculations in order to establish the accuracy and applicability of various approximations for the exchange-correlation energy (XC) functional within DFT.
\par
The paper is arranged as follows: In Section~\ref{sec:details} we will present the necessary details of our calculations, in Section~\ref{sec:results} we will present our results and in Section~\ref{sec:conclusions} we will summarize our results and draw conclusions.
\section{Details of the calculations}\label{sec:details}
\begin{figure}[t]
\includegraphics[width=9cm]{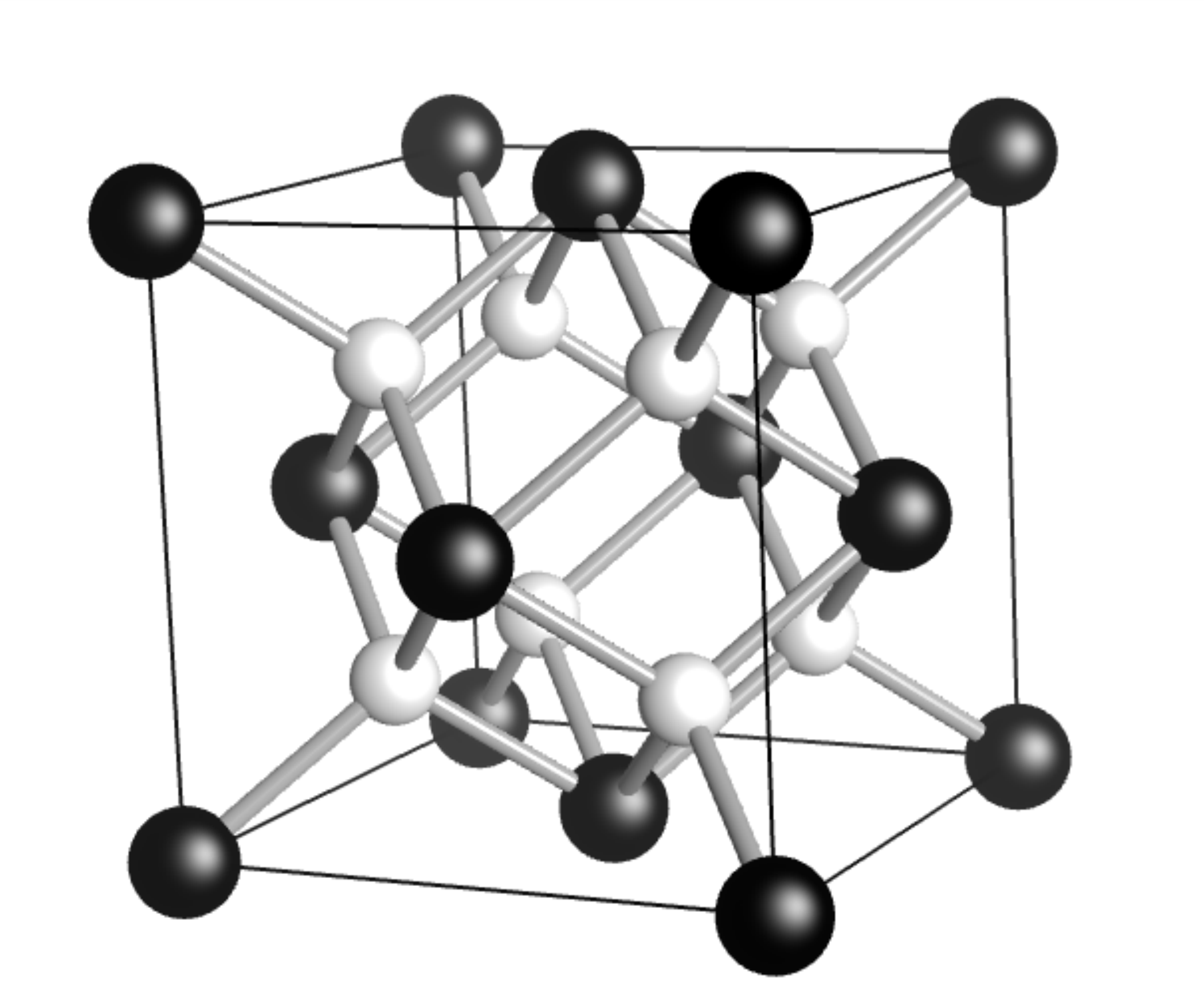}
\caption{\label{fig:structure}The crystal structure of fluorite Cu$_{2}$Se where Cu atoms are depicted by small (white) spheres and Se by large (black) spheres. The Cu-Se bonds are shown in grey.}
\end{figure}

We have performed density functional theory calculations for Cu$_{2}$Se in the fluorite structure, space group Fm$\bar{3}$m, shown in Fig.~\ref{fig:structure}. This structure is an idealized version of the high temperature phase of Cu$_{2}$Se and serves as a good model for determining the accuracy of various approximations for the XC functional. As is shown in Fig.~\ref{fig:structure}, the fluorite structure has the Se atoms positioned at a regular fcc crystal lattice while the Cu atoms occupy the tetrahedral interstitial positions in the fcc lattice.  
\par
The Kohn-Sham equation has been solved using the projector augmented wave method\cite{Blochl} as it is implemented in the Vienna {\it ab initio} simulation package (VASP).\cite{KresseandFurth,KresseandJoubert} Calculations have been performed using the local density approximation (LDA),\cite{Perdew1981} the generalized gradient approximation of Perdew, Burke and Ernzerhof (PBE)\cite{PBE} as well as with the AM05 functional of Armiento and Mattsson.\cite{AM05,Mattsson2008,Mattsson2009} Since local and semi-local approximations to the XC functional are known to perform badly when determining the band gaps of semiconductors and insulators we have in addition performed calculations with the more advanced hybrid functionals of Perdew, Ernzerhof and Burke (PBE0)\cite{PBE0} and of Heyd, Scuseria and Ernzerhof (HSE).\cite{Heyd2003,Heyd2006} These functionals have been found to be an improvement of the LDA and PBE when regarding for example bond distances and dissociation energies in molecules as well as structures and band gaps of bulk materials.\cite{Heyd2003,Heyd2006,Paier2005,Matsushita2011} Interestingly, the AM05 has been found to yield the same level of accuracy as the hybrid functionals when regarding bulk properties of crystals, such as lattice constants and bulk modulus.\cite{Mattsson2008}
\par
The PBE0 and HSE are non-local hybrid theories where the exchange part of the exchange-correlation functional includes some non-local contribution. In the PBE0 approximation the XC energy can be expressed by
\begin{equation}\label{eq:PBE0}
E_{xc}^{PBE0} = \frac{1}{4} E_{x}^{HF} + \frac{3}{4} E_{x}^{PBE} + E_{c}^{PBE},
\end{equation}
where $E_{x}^{PBE}$ and $E_{c}^{PBE}$ are the exchange and correlation terms in the PBE approximation and $E_{x}^{HF}$ is the Hartree-Fock exchange energy. In the HSE approximation the XC energy is given by\cite{Heyd2003,Heyd2006} 
\begin{equation}\label{eq:hybrid}
E_{xc}^{HSE} = E_{x}^{HSE} + E_{c}^{HSE},
\end{equation}
where
\begin{eqnarray}
E_{x}^{HSE} &= \alpha E_{x}^{HF,SR}(\omega) + (1-\alpha) E_{x}^{PBE,SR}(\omega)\nonumber \\ 
& + E_{x}^{PBE,LR}(\omega),
\end{eqnarray}
where $LR$ and $SR$ denote long range and short range parts respectively, $\alpha$ is a mixing parameter governing the strength of the non-local exchange and $\omega$ is a screening parameter that controls the spatial range over which the non-local exchange part is important. The correlation energy in the HSE approximation, $E_{c}^{HSE}$, is taken to be identical to the PBE correlation energy as in the PBE0 approximation. The amount of Hartree-Fock exchange in the HSE is the same as in the PBE0, i.e. $\alpha=1/4$, while the range separation parameter, $\omega$, has to be determined by comparison with experimental data. It has been found that $\omega=0.2-0.3$~\AA$^{-1}$ give good results regarding structural as well as electronic properties of materials, with $\omega=0.2$~\AA$^{-1}$ being the optimum choice.\cite{Heyd2003,Heyd2006,Matsushita2011} Unless otherwise specified, $\omega=0.2$~\AA$^{-1}$ has been used in the HSE calculations. We note that for $\omega=0$ Eq.~(\ref{eq:hybrid}) is equivalent to the PBE0 and, furthermore, Eq.~(\ref{eq:hybrid}) asymptotically reaches the PBE for $\omega\rightarrow\infty$.\cite{Heyd2003} We will return to this point in Sec.~\ref{sec:results}.
\par
An issue with the use of common density functional approximations is the description of strongly localized orbitals such as d- and f-orbitals. For this reason we have also investigated how the localization of the Cu 3d-states affects the electronic structure as well as the band gap in the system by employing the use of an on-site Coulomb interaction within the LDA+$U$ approximation, using the approach of Dudarev et al.\cite{Dudarev} We note that in the Dudarev et al. approach the on-site Coulomb interaction, $U$, and on-site exchange interaction, $J$, do not enter separately but only in the combination of an "effective $U$", $U_{eff}=U-J$. However, throughout the presentation we will simply refer to the "effective $U$" as $U$. The $U$ has been applied solely to the Cu 3d-states.
\par
For the main part of the calculations, the plane wave basis set was cut-off at 1000~eV and we have used a k-point mesh of 30$\times$30$\times$30\cite{MonkhorstandPack} for the primitive cell of Cu$_{2}$Se. However, since the hybrid functionals are much more computationally cumbersome than the local and semi-local approximations, we have used a plane wave cut-off of 800~eV and a k-mesh of 10$\times$10$\times$10 in combination with the hybrid functionals.
\begin{figure}[t]
\includegraphics[width=6cm]{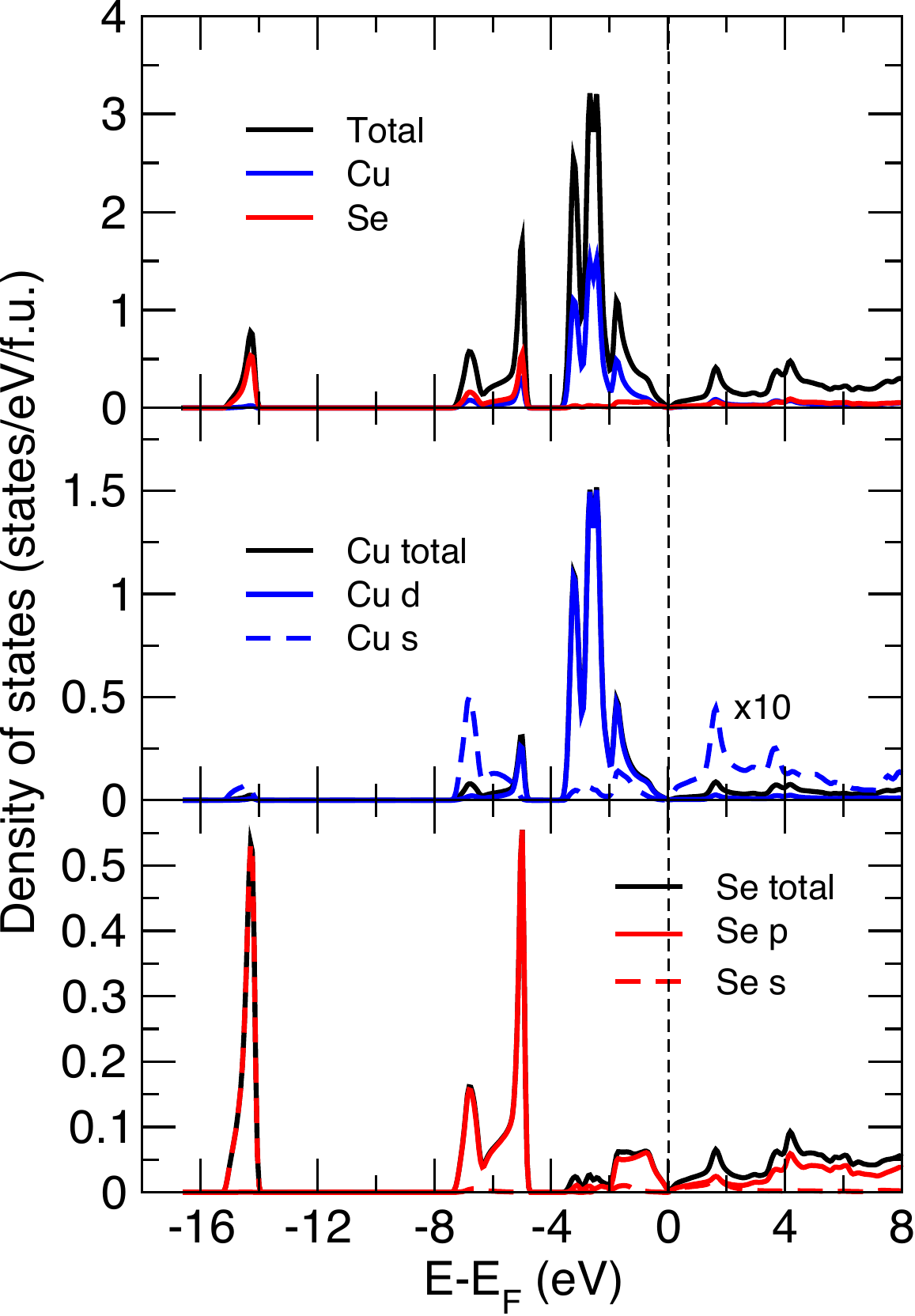}
\caption{\label{fig:dosPBE} (Color online) Calculated density of states and projected density of states obtained by using the PBE. The upper panel is the total DOS and the DOS projected on the Cu and Se atoms. The middle panels is the projected Cu DOS with Cu 3d- and 4s-states marked. The lower panel is the projected Se DOS with Se 4p- and 4s-states marked. The vertical dashed line marks the position of the Fermi level $E_{F}$. In the middle panel the DOS for the Cu 4s-states has been magnified 10 times.}
\end{figure}
\par
The pseudopotential (PP) for Cu has the 3d and 4s electron states treated in the valence while the semi-core 3s and 3p states have been placed in the core of the PP. For Se the PP has the 4s and 4p electron states treated as valence states.

\begin{table}[t]
\caption{\label{tab:lattice_constants} Calculated bulk properties of fluorite Cu$_{2}$Se obtained using the LDA, PBE, AM05, PBE0 and HSE approximations for the XC functional. $a$ is the lattice constant and $B$ is the bulk modulus. The lattice constants and bulk modulus are evaluated using a Birch-Murnaghan 3rd order equation of state.\cite{Birch} The lattice constants and bulk modulus in the second and third columns correspond to the XC functional in the first column while the fifth and sixth columns correspond to the $U$ given in the fourth column within the LDA+$U$ approximation.}
\begin{ruledtabular}
\begin{tabular}{cccccc}
XC functional  &  $a$ (\AA) & $B$ (GPa) & $U$ (eV) & $a$ (\AA) & $B$ (GPa) \\
  \hline
LDA & 5.661& 111 & \\
PBE & 5.844 & 82 &  2 &  5.650 & 108 \\
PBE0 & 5.833 & 81 &  4 &  5.637 & 105\\
HSE & 5.838 & 80 & 6 & 5.623 & 103\\
AM05 & 5.722 & 97 & 8 & 5.605 &  100 \\
  & & &  10 & 5.583 &  98\\
   & & & 12 & 5.554 & 97\\
Exp. & 5.759\cite{Heyding}  & &   \\
 \end{tabular}
\end{ruledtabular}
\end{table}

\section{Results}\label{sec:results}
The lattice constants and bulk modulus obtained using the various functionals are summarized in Table~\ref{tab:lattice_constants}. We note that the PBE, PBE0 and HSE functionals give very similar results in good agreement with experimental observations, whereas the LDA overbinds. The AM05 is the functional that gives the lattice constant closest to the experiment with a deviation that is less than 1~\%. When adding the on-site Coulomb interaction, $U$, to the LDA we find that the lattice constants decreases with increased $U$. The bulk modulus also becomes more similar to the PBE, PBE0 and HSE results as $U$ increases. Usually, the application of an on-site interaction is expected to increase the lattice constant since the $U$ reduces the kinetic energy of the electrons by allowing them to lower their energy by localization. This leads to a lessened binding energy for the electrons for which a $U$ has been applied and in turn an increase in the lattice constant. A reduced lattice constant may nevertheless appear since the population of strongly bonding s- or p-states may increase at the expense of the d-state occupation, resulting in an increased binding strength.
\par
\begin{figure*}[t]
\includegraphics[width=16cm]{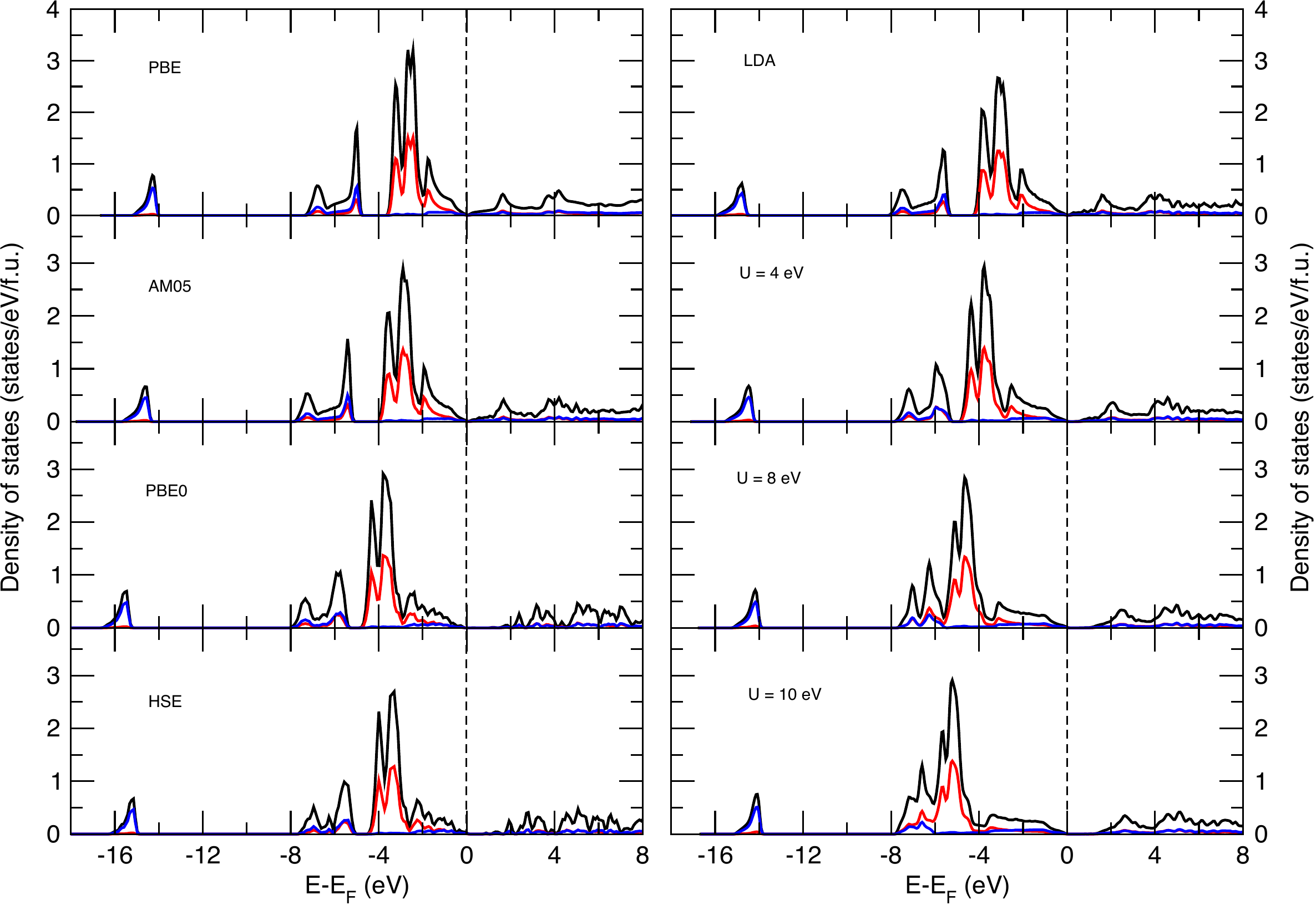}
\caption{\label{fig:dos_all} (Color online) Calculated density of states for PBE, AM05, PBE0 and HSE in the left column and for LDA and LDA+U, with $U=4$~eV, 8~eV and 10~eV, in the right column. The vertical dashed lines marks the position of the Fermi level $E_{F}$ for each case. The total DOS is shown in black and we also show the projected DOS on the two different atomic species; Cu in red and Se in blue.}
\end{figure*}
In Fig.~\ref{fig:dosPBE}, we show the electron density of states (DOS) as evaluated using the PBE functional. It is clear from the results in Fig.~\ref{fig:dosPBE}, that the valence band consists of three regions, a lower region between 14 eV and 15 eV below the Fermi level, which solely is built up by Se 4s-states, a middle region in-between 5 eV and 7.5 eV below the Fermi level that is a mixture of states derived from both Cu and Se, and an upper region from the Fermi level down to 3.5~eV that mostly consists of Cu 3d-states even though there are mixing with the Se 4p-states. The Cu 4s-states are also present in all three regions.  We note that there is a Cu 3d peak in the DOS at about 3~eV below the Fermi level that is positioned in-between two regions of Cu 3d and Se 4p hybridized regions, the lower region between about -8~eV to -5~eV and the upper region from about -2~eV to the Fermi level. This behavior of a Cu 3d peak centered in-between regions of hybridized Cu 3d and Se 4p states has been found experimentally by Domashevskaya et al.,\cite{Domashevskaya} even though the exact positions of the peaks differs which we will turn to shortly. We also note that the result presented in Fig.~\ref{fig:dosPBE} is in agreement with the previous tight binding calculation of Garba and Jacobs\cite{Garba} as well as with the LDA result of Kashida~et~al.\cite{Kashida2003} A notable result is that the PBE functional yield a zero gap between the occupied valence bands and the unoccupied conduction bands. As was mentioned in Section~\ref{sec:intro}, optical studies have reported a band gap of 1.23~eV.\cite{Sorokin1966} This is a rather small gap and density functional theory is known to underestimate the band gap of semiconductors and insulators when using local and semi-local approximations. In order to investigate the band gap for this system further we will have to resort to more complex approximations to the exchange-correlation functional.
\par
In Fig.~\ref{fig:dos_all} we compare the DOS obtained using the LDA, PBE, AM05, PBE0 and HSE as well as LDA+$U$. We find that the overall shape of the DOS is very similar to the PBE, which was discussed previously, except for when very large $U$-values have been applied. In the following discussion we divide the valence states into lower (Se 4s-states), middle (mainly Cu 3d- and Se 4p-states) and upper (including the main Cu 3d peak) valence regions. Compared to the PBE the predominant trend is a shift of the states towards lower energies. For the LDA and AM05 this shift is rather small. For the PBE0 and HSE we find that the lower valence region is found between 15 eV and 16 eV below the Fermi level, compared to 14~eV to 15~eV for the PBE. The middle valence region is also shifted downwards with the same amount ($\sim1$~eV). For the upper valence region the lower edge is shifted downwards with about 1~eV for the hybrid functionals. For very large $U$ the upper valence states are shifted more than the states in the middle valence region and for $U=8$~eV and 10~eV the upper and middle valence regions from the PBE description have merged into a single region. We note that the valence band DOS obtained by the PBE0 and the HSE functionals are in very good agreement with measured x-ray photoelectron spectra (XPS).\cite{Kashida2003,Domashevskaya} The main peak in the spectra at 3.5~eV come from Cu 3d-states and is excellently reproduced by the main Cu peak obtained in the DOS from the PBE0 and the HSE approximations as well as for LDA+$U$ with $U=4$~eV, see Fig.~\ref{fig:dos_all}. For larger $U$-values the Cu d-states are found at too low energies. The LDA, PBE and AM05 have the Cu d-states at slightly to high energy compared to the experiment. As regards the conduction bands and especially the band gap, we find that all approximations have a region of rather small DOS above the Fermi level. 
\par
\begin{figure*}[t]
\includegraphics[width=16cm]{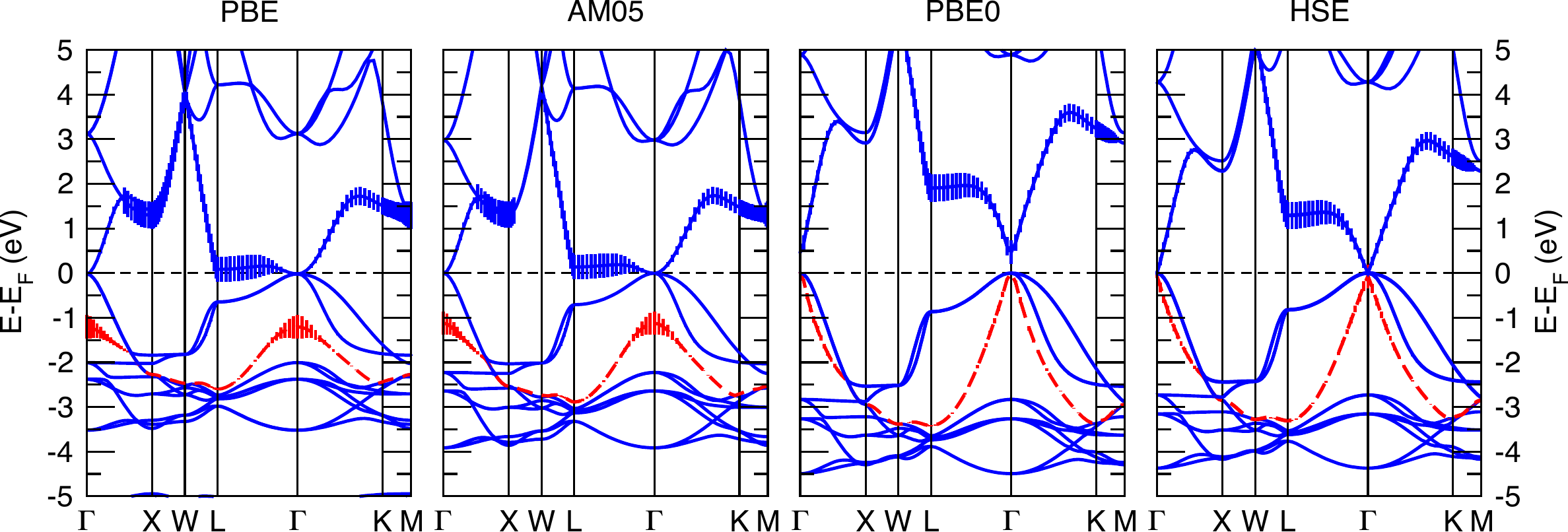}\\
\vspace{0.2cm}
\includegraphics[width=16cm]{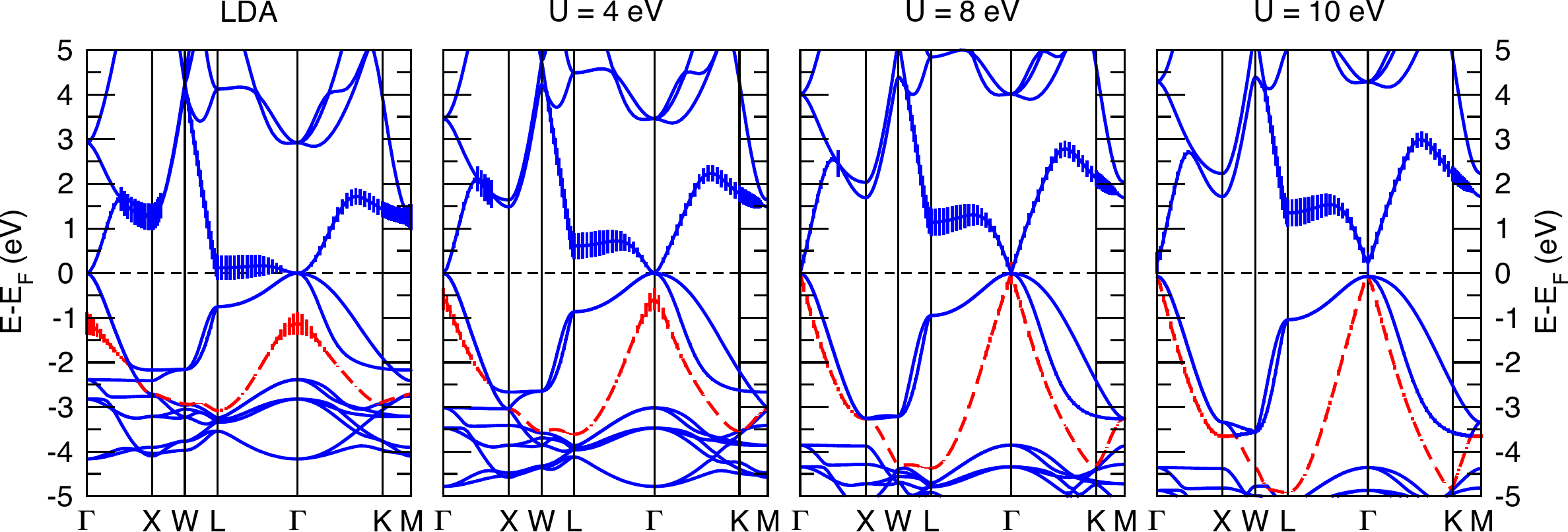}
\caption{\label{fig:PBE+Hybrid} (Color online) Calculated band structures along directions of high symmetry in the Brillouin zone for Cu$_{2}$Se obtained using the LDA, PBE, AM05, PBE0 and HSE functionals.  The vertical solid lines label high symmetry coordinates while the horizontal dashed line marks the position of the Fermi level, $E_{F}$. Note the common feature that in order for a gap to be open the HOMO-2 band, marked by the (red) dashed band, has to reach the Fermi level, which only happen in the case of the PBE0 and LDA+$U$ with $U$=10~eV. The fat bands show the amount of Cu 4s character of the lowest conduction band and the three highest valence bands.}
\end{figure*}
\par
Figure~\ref{fig:PBE+Hybrid} shows the calculated band structure for the upper valence band as well as for a few conduction bands along high symmetry directions in the Brillouin zone. As is clear, only the PBE0 and LDA+$U$ with $U=10$~eV calculations reveal a band gap. In all other cases the highest occupied molecular orbital (HOMO) and the lowest unoccupied molecular orbital (LUMO) touch at the $\Gamma$-point.  The LDA, PBE and AM05, has an almost flat conduction band along the $\Lambda$ line ($\Gamma$ to L), where the LUMO at L is only $\sim0.1$~eV above the Fermi level. For these functionals the lowest conduction band are degenerate with two valence bands at $\Gamma$. The same is also true for the HSE approximation as well as for LDA+$U$ with $U=4$~eV and 8~eV. However, as can be seen in Fig.~\ref{fig:PBE+Hybrid}, the HOMO-2 band at the $\Gamma$-point behaves differently depending on the XC functional. This band is for the PBE band structure found at 1.19~eV below the Fermi level, while for the HSE calculation it is found at 0.07~eV below the Fermi level. For the PBE0 this band has moved up to the Fermi level and the HOMO at the $\Gamma$-point is triply degenerate with a direct gap of 0.47~eV. The same situation is the case for $U=10$~eV but the gap is 0.30~eV. 
\par
When an on-site Coulomb exchange is added to the Cu 3d-states, the main effect is that the Cu 3d-states are pushed further down below the Fermi level as witnessed by the projected DOS in Fig.~\ref{fig:dos_all}. We find that the upper valence bands in Fig.~\ref{fig:PBE+Hybrid} is becoming broader and the conduction bands are shifted towards higher energies. The major effect in the band structure of Fig.~\ref{fig:PBE+Hybrid} is a shift in the low-lying conduction band along the $\Lambda$-line ($\Gamma$~to~L). In order for a gap to be open very large $U$-values are required. The lower right panel of Fig.~\ref{fig:PBE+Hybrid} shows the resulting band structure with $U=10$~eV. The behavior of the valence and conduction bands with $U=10$~eV is rather similar to the PBE0 results in the Fermi level near region, i.e. about -2~eV to 2~eV, with a direct band gap at $\Gamma$ of similar size as the PBE0. We find that the HOMO-2 band is key for the formation of a band gap within the LDA+$U$ approximation. At $U=8$~eV, this band just reaches the Fermi level and marks a transition situation and a small increase of the Coulomb exchange interaction to 8.1~eV (not shown) gives a small gap at $\Gamma$.
\par
\begin{table*}[t]
\caption{\label{tab:Gaps} Calculated transitions between HOMO and LUMO levels at high symmetry points in the Brillouin zone. Energies are given in eV.}
\begin{ruledtabular}
\begin{tabular}{ccccccccc}
$\Delta \varepsilon$ &  PBE  & AM05 & PBE0 & HSE  & LDA & $U=4$~eV & $U=8$~eV & $U=10$~eV \\
  \hline
$\Gamma\rightarrow\Gamma$ & 0.00 & 0.00 & 0.47  & 0.00 & 0.00 & 0.00 & 0.00 & 0.30\\
$\Gamma\rightarrow{\rm L}$ & 0.10 & 0.14 & 1.91 & 1.29  & 0.12 & 0.61 & 1.14 & 1.42\\
$\Gamma\rightarrow{\rm X}$ & 1.31 & 1.31 & 2.91 & 2.28 & 1.26 & 1.49 & 1.69 & 1.78\\
\\
${\rm L}\rightarrow\Gamma$ & 0.63 & 0.70 & 1.33 & 0.82  & 0.75 & 0.86 & 0.95 & 1.27 \\
${\rm L}\rightarrow{\rm L}$ & 0.73 & 0.85 & 2.77 & 2.11  & 0.87 & 1.47 & 2.09 & 2.39\\
${\rm L}\rightarrow{\rm X}$ & 1.94 & 2.02 & 3.78 & 3.11  & 2.01 & 2.35 & 2.64 & 2.76\\
\\
${\rm X}\rightarrow\Gamma$ &  1.82 & 2.03 &  3.00 & 2.44  & 2.17 & 2.66 & 3.25 & 3.57\\
${\rm X}\rightarrow{\rm L}$ & 1.92 & 2.17 & 4.44 & 3.73  & 2.29 & 3.27 & 4.39 & 4.69\\
${\rm X}\rightarrow{\rm X}$ & 3.13 & 3.34 & 5.45 &  4.73  & 3.43 & 4.15 & 4.94 & 5.05\\
 \end{tabular}
\end{ruledtabular}
\end{table*}
 Common features of the band structures shown in Fig.~\ref{fig:PBE+Hybrid} are that the highest valence band, as well as the lowest conduction band, are found along the $\Lambda$ line and that the ordering of the bands is similar irrespective of approximation used. We also note that the valence bands obtained by the PBE0 and HSE are very similar, as is also witnessed by the DOS in Fig.~\ref{fig:dos_all}. The difference between these two hybrid approximations lies in the obtained conduction bands, with PBE0 these bands are found higher above the Fermi level compared to the HSE. The band structure data shown in Fig.~\ref{fig:PBE+Hybrid} are quantified in Table~\ref{tab:Gaps} where we show the evaluated energy differences between HOMO and LUMO levels at various high symmetry points in the Brillouin zone. 
\par
\begin{figure}[b]
\includegraphics[width=8.5cm]{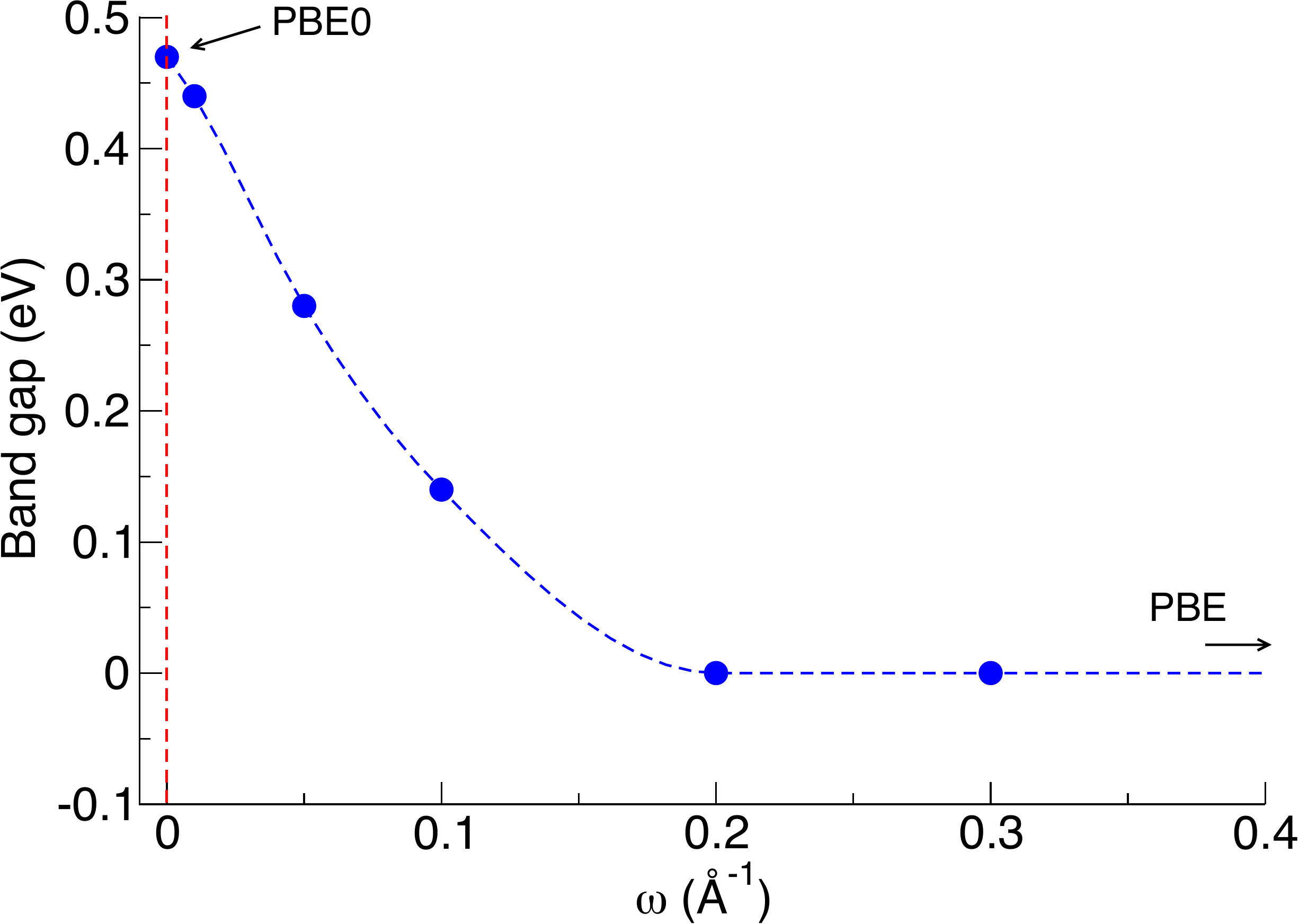}
\caption{\label{fig:omega} (Color online) Evaluated band gaps using various values for the range separation parameter $\omega$ within the HSE approximation. Note that $\omega=0$ in the figure is the PBE0 result. Furthermore, as $\omega\rightarrow\infty$ the PBE result is obtained. The points for $\omega=0.01$, 0.05, 0.1, and 0.3~\AA$^{-1}$ have been evaluated using the lattice constant obtained with  $\omega=0.2$~\AA$^{-1}$ and the points are connected by a fitted spline. The vertical dashed line marks $\omega=0$.}
\end{figure}
\par
We have shown that the existence of a band gap is closely related to the position of the HOMO-2 level at the $\Gamma$-point: If this band reaches the Fermi level a gap is opened. Common for all calculations shown in Fig.~\ref{fig:PBE+Hybrid} is that the triply degenerate state at the Fermi level at $\Gamma$ is derived from hybridization between Cu 3d-states with t$_{2{\rm g}}$ symmetry and Se 4p-states. The HOMO-2 band at $\Gamma$ is derived from hybridized Cu 4s- and Se 4s-states, however, if a gap is open this is no longer the case: Whenever a gap is open the HOMO-2 band reaches the Fermi level and the Cu 4s and Se 4s character is found in the LUMO. This can be seen in Fig.~\ref{fig:PBE+Hybrid} where we show the amount of Cu 4s character of each of the bands close to the Fermi level. As shown in Fig.~\ref{fig:PBE+Hybrid}, there is a region close to $\Gamma$ where the HOMO-2 band has a rather large Cu 4s character in the PBE, AM05 and LDA band structures. In the HSE band structure the Cu 4s character of this band is much more narrow about the $\Gamma$-point, however, it is still present at $\Gamma$. For the PBE0 bands there is no Cu 4s character for this band at $\Gamma$. The same transition of the band character is found for the LDA+$U$ calculations, where $U=8$~eV is a transition state with Cu 4s character remaining in the HOMO-2 at $\Gamma$ which is completely transferred to the LUMO for $U=10$~eV. 
\par
Since the PBE0 and HSE calculations yield very similar valence bands but differ significantly regarding the positions of the conduction bands, it is relevant to analyze the differences between these two approximations. In~Fig.~\ref{fig:omega} we show the calculated band gaps obtained by varying the range separation parameter, $\omega$, within the HSE approximation. The standard values for this parameter are $0.2-0.3$~\AA$^{-1}$, where $\omega=0.2$~\AA$^{-1}$ is considered to be the optimum choice.\cite{Heyd2003,Heyd2006,Matsushita2011} We find that by going below the optimized $\omega$ window, a gap opens up, and the PBE0 value is recovered at $\omega=0$. The reason for this behavior can be linked to that in the HSE approximation, long range exchange contributions are neglected, see Ref.~\onlinecite{Heyd2003}. By writing the PBE0 expression in Eq.~(\ref{eq:PBE0}) in short and long range parts along the lines of the HSE approximation\cite{Heyd2003} we find that
\begin{equation}\label{eq:difference}
E_{xc}^{PBE0}-E_{xc}^{HSE} = \alpha \left(E_{x}^{HF,LR}(\omega)-E_{x}^{PBE,LR}(\omega) \right),
\end{equation}
where $\alpha$ is the amount of Hartree-Fock exchange included (1/4 in the present case). Thus the difference between PBE0 and HSE is the range of the included long-range exchange. By varying $\omega$ it is possible to bring the HSE result in closer agreement with the PBE0, see Fig.~\ref{fig:omega}.  
\par

\section{Discussion and Conclusions}\label{sec:conclusions}
We have performed density functional theory calculations of the electronic structure of fluorite Cu$_{2}$Se using a variety of approximations for the XC energy functional. We find that the PBE, PBE0 and HSE approximations give very similar results regarding lattice constants and bulk modulus. 
The hybrid functionals PBE0 and HSE both show excellent agreement with experimental valence band spectra, although all functionals give rather similar results regarding the overall electronic structure, see Fig.~\ref{fig:dos_all}. In addition, we find that the PBE0 is the only functional that yield a non zero band gap. It is also possible to open a gap by adding an on-site Coulomb interaction to the Cu 3d-states. However, we conclude that localization of the Cu 3d-states cannot account for the formation of a gap since this requires unphysically large $U$-values. Furthermore, compared to experimental valence band spectra such large $U$-values push the Cu 3d-states too low.
\par
We also find that the presence of a band gap is dictated by the relative position of the Cu 4s and Se 4s hybridized band and the Cu 3d and Se 4p hybridized band at the $\Gamma$-point. When a gap is open, i.e. for the PBE0 and for very large $U$, the Cu 4s and Se 4s hybridized band is found in the LUMO above the triply degenerate HOMO that consists of Cu 3d and Se 4p hybridized bands. In all other cases the Cu 4s and Se 4s hybridized band is positioned below the Fermi level. 
\par
In comparison to experimental band gaps we conclude that the present calculations do not reproduce the obtained optical gap of 1.23~eV obtained by Sorokin et al.\cite{Sorokin1966} However, this can be due to several reasons where one is related to the structure of the system. It is possible that defects, most likely Cu vacancies or interstials, affects the electronic structure in such a way that larger gaps are obtained and further investigations along these lines are under way. 
We point out that the experiment was performed at room temperature and for the ordered low temperature phase of Cu$_{2}$Se which has a lower symmetry than the fluorite structure. For Ag$_{2}$Te, which is similar to Cu$_{2}$Se, GGA calculations give a zero gap for the fluorite structure, while a gap opens up for the monoclinic structure.\cite{Zhang2011} We speculate that the same may also be true for Cu$_{2}$Se, i.e. that a sizeable gap opens for the low temperature phase and that the fluorite structure has a vanishingly small gap. 
\par
Furthermore, our calculations reveal that the band structure around the $\Gamma$-point is such that it would not create a clear peak in the optical spectrum but rather an energy independent low intensity signal which can easily be confused with a background. In addition, our DOS calculations shown in Fig.~\ref{fig:dos_all} reveal a region with very small DOS above the Fermi level irrespective of the XC energy functional that has been used.
\par
In addition, it has been found that HSE is better compared to the PBE0 at describing the band gap in small gap semiconductors, where the latter is found to overestimate the size of the band gap.\cite{Matsushita2011} Thus, in summary, despite the fact that PBE0 is the only functional opening up a gap, we believe that HSE with $\omega\sim0.2$~\AA$^{-1}$ is the functional that more correctly describes the electronic structure of Cu$_{2}$Se. In order to further analyze the optical spectra by theory it will be necessary to resort to solving the Bethe-Salpeter equation or using time-dependent DFT where optical gaps can be described properly due to the inclusion of the electron-hole interaction in these methods. 

\section{Acknowledgements}
This work was financed through the EU project NexTec, VR (the Swedish Research Council), and SSF (Swedish Foundation for Strategic Research). The computations were performed on resources provided by the Swedish National Infrastructure for Computing (SNIC) at the National Supercomputer Centre in Link{\"o}ping (NSC).

\end{document}